\documentclass{emulateapj}

\newcommand{\beq}{\begin{equation}}
\newcommand{\beqa}{\begin{eqnarray}}
\newcommand{\eeq}{\end{equation}}
\newcommand{\eeqa}{\end{eqnarray}}

\newcommand{\bfx}{\mathbf{x}} 
\newcommand{\bfk}{\mathbf{k}}

\newcommand{\simgt}{\lower.5ex\hbox{$\; \buildrel > \over \sim \;$}}
\newcommand{\simlt}{\lower.5ex\hbox{$\; \buildrel < \over \sim \;$}}

\newcommand{\mi}{\mathrm{i}}
\newcommand{\mj}{\mathrm{j}}

\shorttitle{Covariance matrix of the matter power spectrum}
\shortauthors{Takahashi et al.}

\begin{document}

\title{Simulations of Baryon Acoustic Oscillations II:
Covariance matrix of the matter power spectrum}

\author{Ryuichi Takahashi\altaffilmark{1},
Naoki Yoshida\altaffilmark{2},
 Masahiro Takada\altaffilmark{2},
Takahiko Matsubara\altaffilmark{1}, 
Naoshi Sugiyama\altaffilmark{1,2}, 
Issha Kayo\altaffilmark{2}, 
Atsushi J. Nishizawa\altaffilmark{3},
Takahiro Nishimichi\altaffilmark{4},
Shun Saito\altaffilmark{4},
Atsushi Taruya\altaffilmark{2,5}}

\affil{\altaffilmark{1} Department of Physics,
 Nagoya University, Chikusa, Nagoya 464-8602, Japan}
\affil{\altaffilmark{2} Institute for Physics and Mathematics of the Universe,
The University of Tokyo, 5-1-5 Kashiwa-no-ha, Kashiwa, Chiba 277-8568, Japan}
\affil{\altaffilmark{3} National Astronomical Observatory of Japan,
 2-21-1 Osawa, Mitaka, Tokyo, 181-8588, Japan}
\affil{\altaffilmark{4} Department of Physics, School of Science,
 The University of Tokyo, Tokyo 113-0033, Japan}
\affil{\altaffilmark{5} Research Center for the Early Universe,
 The University of Tokyo, Tokyo 133-0033, Japan}

\begin{abstract}
We use 5000 cosmological $N$-body simulations of $1h^{-3}$Gpc$^{3}$ box for the
concordance $\Lambda$CDM model in order to study the sampling variances
of nonlinear matter power spectrum.
We show that the non-Gaussian errors can be important
even on large length scales relevant for baryon acoustic oscillations
(BAO). Our findings are (1) the non-Gaussian errors degrade the
cumulative signal-to-noise ratios ($S/N$) for the power spectrum
amplitude by up to a factor of 2 and 4 for redshifts $z=1$ and $0$,
respectively. (2) There is little information on the power spectrum
amplitudes in the quasi-nonlinear regime, confirming the previous results.
(3) The distribution of power spectrum estimators at BAO scales, among
the realizations, is well approximated by a Gaussian distribution with
variance that is given by the diagonal covariance component.
(4) For the redshift-space power spectrum, the
degradation in $S/N$ by non-Gaussian errors is mitigated due to
nonlinear redshift distortions. (5) For an actual galaxy survey, the
additional shot noise contamination compromises the cosmological information
inherent in the galaxy power spectrum, but also mitigates 
the impact of non-Gaussian errors. The $S/N$ is degraded by up to
$30\%$ for a WFMOS-type survey. 
(6) The finite survey volume causes additional non-Gaussian errors via
the correlations of long-wavelength fluctuations with the fluctuations
we want to measure, further degrading the $S/N$ values by about 30\%
even at high redshift $z=3$.
\end{abstract}


\section{Introduction}

Baryon acoustic oscillations (BAO) are imprinted in the 
distribution of galaxies,
of which the characteristic length scale can be used as a standard ruler 
in the universe
 (e.g. Eisenstein, Hu \& Tegmark 1998; Blake \& Glazebrook 2003;
 Seo \& Eisenstein 2003; Matsubara 2004; Guzik, Bernstein \& Smith 2007).
BAO provides a powerful way of probing the nature of dark energy.
Large galaxy surveys such as the Sloan Digital Sky Survey and
 two degree Field Survey detected the BAO signature and
 provided constraints on the dark energy equation of state
 (Cole et al. 2005; Eisenstein et al. 2005; Percival et al. 2007; Okumura
 et al. 2008; Gaztanaga, Cabre \& Hui 2008; Sanchez et al. 2009).
Future larger surveys are aimed at measuring the BAO scale
more accurately and hence yielding tighter constraints 
on the nature of dark energy
(see e.g. Benitez et al. 2008). 

The BAO signature in the galaxy power spectrum is very small,
of the order of a few percent modulation in amplitude,
and hence measurements of the precise length scale
are hampered by a number of effects.
For example, nonlinear gravitational evolution,
redshift space distortion, galaxy formation processes and the
 associated scale-dependent bias,
all compromise a robust detection.
Accurate theoretical models are clearly needed. 
A number of authors resort to 
using numerical simulations (Meiksin, White \& Peacock 1999; Seo \&
 Eisenstein 2005; Huff et al. 2007; Smith, Scoccimarro \& Sheth 2007, 2008;
 Angulo et al. 2008; Takahashi et al. 2008; Seo et al. 2008;
 Nishimichi et al. 2008)
whereas others use perturbation theory (Crocce \& Scoccimarro 2006, 2008;
 Jeong \& Komatsu 2006, 2009; Nishimichi et al. 2007;
 McDonald 2007; Matarrese \& Pietroni 2007, 2008; Pietroni 2008;
 Matsubara 2008a,b; Taruya \& Hiramatsu 2008;
 Takahashi 2008; Nomura, Yamamoto \& Nishimichi 2008; Rassat et al. 2008).
It is important to note that
one needs accurate estimates not only for the power spectrum 
but also its covariance
 (e.g. Scoccimarro, Zaldarriaga \& Hui 1999; Meiksin \& White 1999;
 Habib et al. 2007).
The covariance describes statistical uncertainties of the power spectrum
measurement as well as the band powers at different wavenumbers are
 correlated with each other. Hence once the well-calibrated covariance
is obtained, one can derive unbiased, robust constraints on cosmological
parameter from the measured power spectrum (see Ichiki et al. 2008 for
 such an example to show the importance of the covariance estimation).

The power spectrum covariance matrix has only diagonal elements
for the Gaussian density fluctuations 
 (e.g. Feldman, Kaiser \& Peacock 1994).
The relative error of the power spectrum of a given wavenumber 
is then simply given by the square root of the number 
of Fourier modes 
available from the survey volume. 
However, at small length scales, 
non-vanishing off-diagonal parts of the covariance
arise due to the mode coupling
(Scoccimarro, Zaldarriaga \& Hui 1999; Meiksin \& White 1999; Smith 2008).
This non-Gaussian contribution is described by 
the trispectrum or the Fourier transform of the 4-point correlation
function. Cooray \& Hu (2001) 
used the halo model to estimate the trispectrum contribution
and showed that 
the non-Gaussian errors do degrade the precision of cosmological 
parameter determination, and therefore cannot be ignored for planned 
future surveys
(see also Takada \& Jain 2008; Eifler, Schneider \&
Hartlap 2008).
Also recently, Smith (2008) studied
the covariance matrix of the halo power spectrum using numerical
simulations.
Sefusatti et al. (2006) also studied the power spectrum covariance
 using PTHalos (Scoccimarro \& Sheth 2002).

In this paper, we use an unprecedentedly large number of 
simulation realizations to estimate
the covariance matrix of the matter power spectrum in
both real and redshift space. Our sample is more than  
2 orders of magnitude larger than those used in 
the previous works, yielding well-converged
estimates on the power spectrum covariance.
We compare our simulation results with the analytical estimates based on
perturbation theory and halo model. 
In these comparisons we also include the new effect of non-Gaussian
errors that inevitably arise for a finite-volume survey,
 as first pointed out in Rimes \& Hamilton
(2005; also see Rimes \& Hamilton 2006; Hamilton, Rimes \& Scoccimarro
2006; Neyrinck, Szapudi \& Rimes 2006; Neyrinck \& Szapudi 2007, and Lee
\& Pen 2008 for the observational implication based on the SDSS data).
By using this large number of the realizations, we also study how the power
spectrum estimates are distributed in different realizations, i.e. the
probability distribution of power spectrum, 
and then compute the higher-order moments, skewness and kurtosis, to
examine the overall impact of power spectra at high sigma ends. 
Furthermore 
we estimate the expected signal-to-noise ratio for measuring the
power spectrum for a future galaxy survey, taking into account the shot
noise contamination and the non-Gaussian errors. 

Throughout the present paper, we adopt the standard $\Lambda$CDM model
with matter density $\Omega_{m} =0.238$, baryon density
$\Omega_{\rm b}=0.041$, cosmological constant $\Omega_\Lambda=0.762$,
spectral index $n_{\rm s}=0.958$, amplitude of fluctuations $\sigma_8=0.76$, 
and expansion rate at the present time $H_{0}=73.2$km s$^{-1}$
Mpc$^{-1}$,
which are consistent with the WMAP 3-year results (Spergel et al. 2007).

\section{Numerical Simulations}
\label{sec:sim}

We use the cosmological simulation code Gadget-2
 (Springel, Yoshida \& White 2001; Springel 2005).
We employ $256^3$ particles in a volume of $1000h^{-1}$ Mpc on a side.
We generate initial conditions 
of the seed density perturbations
 at $z=20$ based on the standard Zel'dovich approximation
 using the matter transfer function calculated by CAMB (Code for Anisotropies
 in the Microwave Background: Lewis, Challinor \& Lasenby 2000).
We ran 5000 realizations of Particle Mesh (PM) simulations for the
fiducial cosmological model, and use the snapshot outputs at $z=3, 1$
and $0$ to study the power spectrum covariances. 

To calculate the Fourier transform of 
the density field, 
denoted as $\tilde{\delta}(\bfk)$, 
we first assign the
$N$-body particles onto $N_{\rm grid}^3=512^3$ grids based on the
 cloud-in-cell method
and then perform FFT\footnote{FFTW home page: http://www.fftw.org/}.
We also correct the effect of the cloud-in-cell assignment scheme as
 $\tilde{\delta}(\bfk) \rightarrow \tilde{\delta}(\bfk)$
 $\times$ $[{\rm sinc}(k_x L/2 N_{\rm grid})$
 ${\rm sinc}(k_y L/2 N_{\rm grid})$
 ${\rm sinc}(k_z L/2 N_{\rm grid})]^{-2}$ with ${\rm sinc}(x)=\sin x/x$
 (Hockney \& Eastwood 1988; Angulo et al. 2008). 
The binned power spectrum for a given realization is
estimated as 
\beq
  \hat{P}(k)=\frac{1}{N_k} 
\sum_{|\bfk|\in k}
 \left| \tilde{\delta}(\bfk) \right|^2,
\label{eq_pk}
\eeq
where the summation runs over all the Fourier modes whose length is in
the range $k-\Delta k/2\le |\bfk|\le k+\Delta k/2$ for a given bin width
$\Delta k$.
Here $N_k$ is the number of modes taken for the summation 
and is given as $N_k=\sum_{|\bfk|\in k}\approx 
4\pi k^2\Delta k/(2\pi/L)^3=Vk^2\Delta k/(2\pi^2)$ for the limit 
$k\gg 1/L$, where 
$L$ is the simulation box size and $V$ is the volume given by $V=L^3$.
The shot noise is not subtracted, since this effect is very small. 
The ensemble average of the power spectrum estimator is then computed by
averaging the estimated spectra over the realizations: 
$P(k)=\langle \hat{P}(k) \rangle$.

We have checked that 
our simulation result for the power spectrum agrees with 
the higher resolution TreePM result within $1\%(3\%)$ at
 $k<0.2(0.4)h$/Mpc \footnote{The agreement is achieved in real
 space. In redshift space, PM simulations somewhat underestimate the
 power spectrum by $20 (10) \%$ at $z=0,1 (3)$
 at small length scales ($k = 0.4h$/Mpc).}
 (here the Nyquist wavenumber is $k=0.8h/$Mpc).
If the initial redshift is set to be higher, e.g. $z=50$,
the results agree within $2\%$ for $k<0.2h$/Mpc and $10\%$ for
$k<0.4h$/Mpc for $z=0,1,3$.
This is sufficient for our purpose, which is to estimate the impact of
nonlinear clustering on the power spectrum covariances at BAO scales. 

\section{Covariance Matrix}

The covariance between the power spectra, $P(k_1)$ and $P(k_2)$, is 
estimated from the simulation realizations and can be
formally expressed in terms of the Gaussian and non-Gaussian
contributions (e.g. Scoccimarro, Zaldarriaga \& Hui 1999;
 Meiksin \& White 1999): 
\beqa
  {\rm cov}(k_1,k_2) \equiv
  \left< \left( \hat{P}(k_1) - P(k_1) \right) \left( \hat{P}(k_2) - P(k_2)
 \right) \right>  \nonumber \\
 =  \frac{2}{N_{k_1}} P^2(k_1) \delta^K_{k_1,k_2} \hspace{3.8cm}  \nonumber \\
 + \frac{1}{V} \int_{|\bfk{}'_1|\in k_1} \int_{|\bfk{}'_2|\in k_2}
 \!\!\frac{d^3 \bfk_1'}{V_{k_1}} \frac{d^3 \bfk_2'}{V_{k_2}}
 ~T(\bfk_1',-\bfk_1',\bfk_2',-\bfk_2'),
\label{cov}
\eeqa
where $T$ is the trispectrum, 
and
$\delta^K_{k_1k_2}$ is the Kronecker-type delta function defined
such that $\delta^K_{k_1k_2}=1$ if $k_1=k_2$ within the bin width,
otherwise zero. 
The integration range in the second term is, as in Eq.~(1), 
confined to the Fourier
modes lying in the range $k_1-\Delta k/2\le k
\le k_1+\Delta k/2$, and $V_{k_i}$ ($i=1,2$) denotes the integration
volume in Fourier space given by $V_{k_1}\approx 4\pi k^2\Delta k$
for the case of $k\gg\Delta k$.

The first term of the covariance matrix represents the Gaussian error 
contribution ensuring that the two power spectra of different
wavenumbers are uncorrelated, while the second term gives the
non-Gaussian errors that include correlations between power spectra at
different $k$'s arising from nonlinear mode coupling. 
Both the terms scale with the simulation box volume as $\propto
1/V$. It should be also noted that the non-Gaussian term does not depend on
the bin width (because $\int_{|\bfk'|\in k}d^3\bfk'/V_k\approx 1$),
 so increasing $\Delta k$ only 
reduces 
the Gaussian contribution via the dependence $N_k\propto \Delta k$. 
However, the cumulative 
signal-to-noise ratio we will study below is independent of
the assumed
$\Delta k$. 

We will compare the simulation results 
with two analytical approaches
to estimate the covariance matrix: 
(1) perturbation theory and (2) halo model.
In perturbation theory, 
following Scoccimarro et al. (1999; also see Neyrinck \& Szapudi 2008), 
the power spectrum and trispectrum are, self-consistently including up to
the third order perturbations of $\tilde{\delta}$, expressed as
\beqa
  P^2(k_1)=P^2_{\rm lin}(k_1) +  2 P_{\rm lin}(k_1) \left[ P_{22}(k_1)
 + P_{13}(k_1) \right], \hspace{0.5cm} \nonumber \\
  T(\bfk_1,-\bfk_1,\bfk_2,-\bfk_2)=12 ~P_{\rm lin}(k_1) P_{\rm lin}(k_2)
 \hspace{1.8cm} \nonumber \\
 \times \left[  F_3(\bfk_1,-\bfk_1,\bfk_2) P_{\rm lin}(k_1)
 + (k_1 \leftrightarrow k_2) \right] \hspace{1.75cm}  \nonumber \\
 + 8 P_{\rm lin}(|\bfk_1-\bfk_2|) \left[ F_2(\bfk_1-\bfk_2,\bfk_2)
 P_{\rm lin}(k_2)  \right. \hspace{1.35cm} \nonumber  \\
 \left. + (k_1 \leftrightarrow k_2) \right]^2, \hspace{5.6cm} 
\label{eq:pt}
\eeqa
where $P_{\rm lin}$ denotes the linear-order spectrum, and 
$P_{22}$ and $P_{13}$ are the one-loop
 corrections 
to the nonlinear power spectrum
 (Makino, Sasaki \& Suto 1992; Jain \& Bertschinger 1994)
 and $F_{2}$ and $F_3$ are the kernels for the second
 and third order density perturbations (e.g. Bernardeau et al. 2002).

In the halo model, the power spectrum is given by a sum of two terms,
the so-called one-halo term and two-halo term (Seljak 2000; 
Ma \& Fry 2000; Peacock \& Smith 2000;
also see Cooray \& Sheth 2002 for a review).
Similarly, the trispectrum consists of four terms, 
 from one to four halo terms:
\begin{equation}
T=T^{\rm 1h}+T^{\rm 2h}+T^{\rm 3h}+T^{\rm 4h}. 
\end{equation}
The explicit expressions of each term can be found in Cooray \& Hu
(2001). 
In nonlinear regime, i.e. large $k$, the 1-halo term gives
dominant contribution to the total power of trispectrum, 
while the different halos
terms become more significant with decreasing $k$ and the 4-halo term
that includes the PT trispectrum contribution becomes
dominant on very small $k$.

To complete the halo model approach, we need suitable models for the
three ingredients: 
the halo mass function (Sheth \& Tormen 1999), 
the halo bias parameters (Mo \& White 1996; Mo,
 Jing \& White 1997), and
the halo mass density profile (Navarro, Frenk \& White 1997), 
 each of which is specified by
halo mass $m$ and redshift $z$ for a given cosmological model.
The details of our halo model implementation
can be found in Takada \& Jain (2003; 2008). 

In our previous paper (Takahashi et al. 2008), we found that a finite
size simulation causes the growth of large-scale density  
perturbations to
be deviated from the linear theory prediction,
and the deviation is well
described by the nonlinear mode coupling. We proposed a method to
``correct'' the deviation in the finite size simulations to obtain  
the ensemble
averaged expectation of power spectrum without running ideal  
simulations with infinite volume
(practically very large volume), as also demonstrated
in Nishimichi et al. (2008). In this paper,
 where we are discussing the power spectrum covariance for
some finite survey volume, we do not have to ``correct'' the  
deviation in
power spectrum of each realization,
because the scatters are already included in the covariance formula
(the terms with $F_2$ in Eq.~[\ref{eq:pt}]).

We use $5000$ realizations of each output redshift 
to directly estimate the covariance matrix
according to Eq.~(\ref{cov}). To be more explicit, denoting 
the power spectrum of the $i$-th realization 
as $\hat{P}_i(k)$,  
we can estimate the covariance as 
\begin{equation}
{\rm cov}(k_1,k_2)= \frac{1}{N_{\rm r}-1}
\sum_{i=1}^{N_{\rm r}} [\hat{P}_i(k_1)-\bar{P}(k_1)]
[\hat{P}_i(k_2)-\bar{P}(k_2)], 
\label{eq:covs}
\end{equation}
where $N_{\rm r}$ is the number of realizations, i.e.  $N_{\rm r}=5000$
in our case, and $\bar{P}(k)$ denotes the mean spectrum computed as
$\bar{P}(k) = (1/N_{\rm r}) \sum_i \hat{P}_i(k)$.
As shown in Appendix,
the accuracy in estimating the covariances scales with the
number of realizations used\footnote{More precisely, the accuracy of
estimating the covariance is determined by the covariance of the power
spectrum covariance that includes up to the 8-point correlation
functions (Kayo et al. in preparation).}. For example, 
the relative accuracy of estimating the diagonal 
covariance elements is found to scale approximately as 
 $(N_{\rm r}/2)^{-1/2}$. Hence, with the aid
of 5000 realizations, we can achieve a few \%-level accuracies in
estimating each elements of the covariance,
an improvement by an order of magnitude over previous works.

\section{Comparison with Theoretical Models}

\subsection{Results in Real Space}

\begin{figure}
\plotone{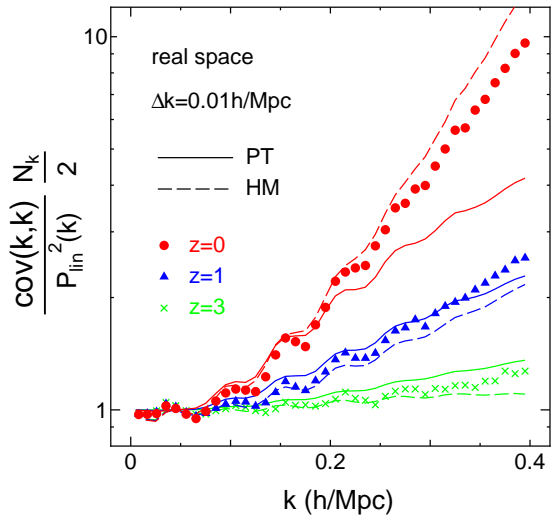}
\caption{The diagonal components of the
power spectrum covariance as a function of wavenumbers, for redshifts
$z=0$, 1 and $3$.  The results are divided by the Gaussian covariance of
the linearly evolving power spectrum. Therefore the deviations from
unity arise from both the nonlinear clustering and the non-Gaussian
errors.
The symbols are the simulation results, while the solid curves show the
results obtained when the perturbation theory (PT) is used to compute the
non-Gaussian covariance.
The dashed curves show the halo model results.
}
\label{pk_disp} \hspace{0.5cm}
\end{figure}

Fig.\ref{pk_disp} shows the diagonal elements of the covariance matrix
as a function of wavenumbers. The diagonal elements plotted are divided
by the Gaussian covariances of 
linear power spectra at each
redshift such that the values become unity in the linear regime limit
($k\rightarrow 0$). 
Therefore
the deviations from unity arise from the non-linear
 evolution of $P(k)$ and the non-Gaussian covariance contribution. 
The cross, triangle and circle symbols show the 
simulation results for redshifts $z=3,$ 1 and 0, respectively. 
Note that we 
adopt the bin width of $\Delta
k=0.01h$Mpc$^{-1 }$ throughout this paper.
The deviations from the Gaussian errors become more significant at lower
redshifts. For comparison, the solid curves show the analytical
predictions obtained when the perturbation theory is employed to estimate
the covariances as described around Eq.~(\ref{eq:pt}). The perturbation
theory  fairly well reproduces the simulation results 
within $20\%$ up to $k<0.24h/$Mpc at $z=0$ and $k<0.4h/$Mpc at $z=1$ and
$3$, respectively.  However, at lowest redshift $z=0$, stronger
nonlinear effects are seen even on these large length scales
corresponding to the BAO scales. 
The dashed curves show 
the halo model results
 which
take
into account of this nonlinear effect.
The halo
model fairly well
 fits the simulation results over the range of
 wavenumbers studied.
At $z=1,3$ the PT predicts the larger variance than the halo model,
 because the one-loop power spectrum in the Gaussian term (\ref{eq:pt})
 overestimate the power spectrum.

\begin{figure*}
\plotone{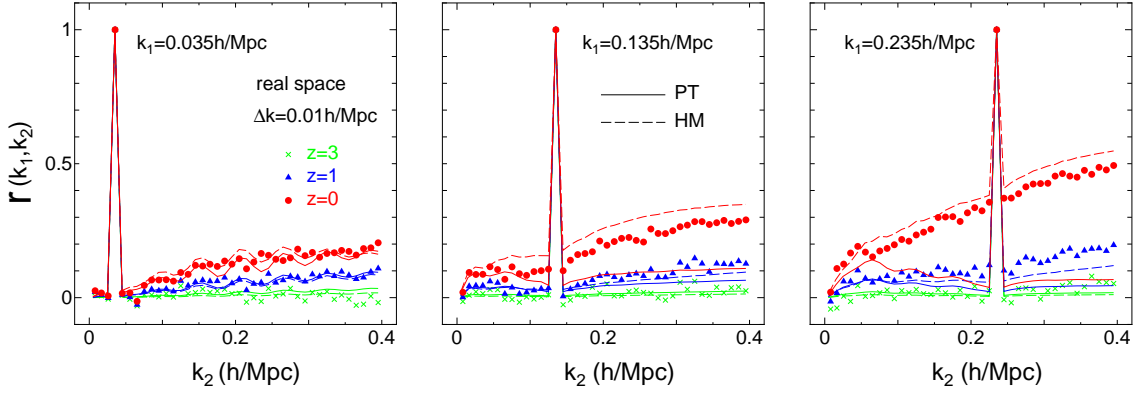}
\caption{The correlation coefficient matrix $r(k_1,k_2)$, defined 
in Eq.~(\ref{cov_norm}), as function of $k_2$, where $k_1$ is kept fixed
to $k_1=0.035$, 0.135 and 0.235$h/$Mpc in the left, middle and right
 panels, respectively. 
The solid curves denote the PT predictions, while the dashed curves
 show the halo model results.
The simulation results at $z=0$ and 1 show greater amplitudes in 
 the off-diagonal covariances than the PT predictions.
}
\label{corrmat}
\end{figure*}

Fig.\ref{corrmat} shows the off-diagonal elements of the covariance
matrix.
For illustrative purpose we study the correlation coefficient matrix 
defined as
\beq 
  r(k_1,k_2)=\frac{{\rm cov}(k_1,k_2)}{\sqrt{{\rm cov}(k_1,k_1)
 {\rm cov}(k_2,k_2)}}.
\label{cov_norm}
\eeq 
The coefficients are normalized so that $r=1$ for the diagonal
components with $k_1=k_2$. For the off-diagonal components $r\rightarrow
1$ implies strong correlation between the two spectra, while $r=0$
corresponds to no correlation. 
Note again that the matrix elements $r$ depend on the bin width: a finer
binning, i.e. a smaller $\Delta k$, decreases the off-diagonal
components. 
First of all, comparing the three panels of Fig.~\ref{corrmat} manifests
that the off-diagonal components have greater amplitudes with increasing
$k$. 
The PT fits the data within $0.1$ for $k<0.15h/$Mpc at $z=0$,
 for $k<0.24h/$Mpc at $z=1$ and for $k<0.29h/$Mpc at $z=3$.
For the redshift dependence, there is almost no cross-correlations
at redshift $z=3$, while there are increasing cross-correlations at lower
redshifts. 

The PT results start to underestimate
the correlation strengths with increasing $k$ and at lower redshifts due
to the stronger nonlinearities. 
Compared to Fig.~\ref{pk_disp},
the PT results are found to be less accurate to describe
the off-diagonal components at $z=1$ and 0. 
The dashed curves are the halo model results, which are in a good
agreement with the simulation results,  especially at $z=0$.
We found that an inclusion of the 2- and 3-halo terms is important to
describe the scale dependences of the off-diagonal correlations. 
However, the halo model displays a sizable disagreement at some scales,
and is not well accurate. 
Therefore a further refinement of the model predictions based
on 
this kinds of large-scale simulations
is needed 
to accurately model the measurement errors of power
spectrum, especially for future high-precision surveys.

\subsection{Results in Redshift Space}

In this section, 
we examine the covariance of the redshift-space power spectrum that is
a more direct observable in galaxy surveys. 
The redshift-space power spectrum in each realization is computed as
follows. Assuming the distant observer approximation, we first calculate
the density perturbations in redshift space as described in
\S~\ref{sec:sim}, but properly taking into account modulations of
$N$-body particle positions in redshift space due to the peculiar
velocities. The density perturbation field is thus given as a function
of wavenumbers $k_\parallel$ and $\bfk_\bot$ that are parallel and
perpendicular to the line-of-sight (taken from one direction in the
simulation box).  As a result, the redshift-space power spectrum $P_s$
is given as a two-dimensional function due to the
statistical isotropy: $P_s(k_\bot,k_\parallel)$. In this paper, for
simplicity, we focus on the spherically averaged redshift-space spectrum
over the shell of a radius $k$ with the width $\Delta k$:
\begin{equation}
\hat{P}_{s0}(k)\equiv \frac{1}{N_k}\sum_{|\bfk'|\in k}|
\tilde{\delta}_s(k'_\parallel,\bfk_\bot')|^2,
\label{eq:ps0s}
\end{equation}
where $k'=\sqrt{k_\parallel^{\prime2}+|\bfk_\bot^{\prime }|^2}$ and $N_k$ is
the number of modes in the spherical shell in redshift space. Likewise,
the covariance matrix of $P_{s0}$ can be estimated by averaging the
spectrum estimators among the simulation realizations as in
Eq.~(\ref{eq:covs}).

According to the linear perturbation theory of structure formation, the
redshift-space power spectrum can be simply related to the real-space
spectrum under the distant observer approximation as 
$P_s(k_\parallel,k_\bot)= ( 1+f \mu^2 )^2 P(k)$, where
$\mu=k_\parallel/k$ is the cosine between the line-of-sight and the
wavevector and $f$ is the linear redshift distortion, expressed in
terms of the linear growth rate $D_1$ as $f=(d \ln D_1/d \ln a)/b$
 (Kaiser 1987) with the bias parameter $b=1$ for the dark matter
 power spectrum.
Note that all the spectra we have considered are for the
total matter distribution. 
Averaging the redshift-space spectrum over the cosine angle $\mu$ yields
the linear theory prediction that is to be compared with
the simulation result given by
Eq.~(\ref{eq:ps0s}):
\begin{equation}
P_{s0}(k) = [ 1+(2/3) f +(1/5) f^2 ] ~P(k).
\label{eq:ps0}
\end{equation}

The prefactor in front of $P(k)$ on the r.h.s. of Eq.~(\ref{eq:ps0})
does not depend on wavevector. Hence, from Eqs.~(\ref{cov}) and
(\ref{eq:ps0}), the linear theory tells that the covariance of the
redshift-space power spectrum (\ref{eq:ps0}) can be simply expressed 
as\footnote{The angular average of the covariance is proportional to
 $\int d\mu (1+f\mu^2)^4$, while the square of the angular averaged
 $P_s(k)$ is proportional to $[\int d\mu (1+f\mu^2)^2]^2$.
Hence the two quantities are not same and the extra factor
in Eq.(\ref{disp_redshift2}) appears.}
\begin{eqnarray}
{\rm cov}_s(k_1,k_2)&\equiv&
\left< 
\left( \hat{P}_{s0}(k_1) - P_{s0}(k_1) \right)
\left( \hat{P}_{s0}(k_2) - P_{s0}(k_2) \right)
\right> 
\nonumber\\
&&\hspace{-5em}=
 \frac{2}{N_{k_1}} P^2_{s0}(k_1)\delta^K_{k_1k_2}  
\frac{1+\frac{4}{3}f + \frac{6}{5} f^2 + \frac{4}{7}
 f^3 + \frac{1}{9}f^4}{\left( 1+\frac{2}{3} f+\frac{1}{5}
 f^2 \right)^2}.
\label{disp_redshift2}
\end{eqnarray}
Due to the additional factor that depends solely on $f$, the
covariance amplitude of $P_{s0}$ is greater than the standard Gaussian
error, $(2/N_k)P_{s0}^2$, by 6, 16 and $20\%$ at $z=0$, 1 and $3$
for the $\Lambda$CDM model, respectively. Note that the covariance form
(\ref{disp_redshift2}) is valid only for the asymptotic limit of large
length scales, and in general the nonlinear clustering effects cause
deviations from the Kaiser formula on the BAO scales (e.g., Scoccimarro
2004).

\begin{figure*}
\plotone{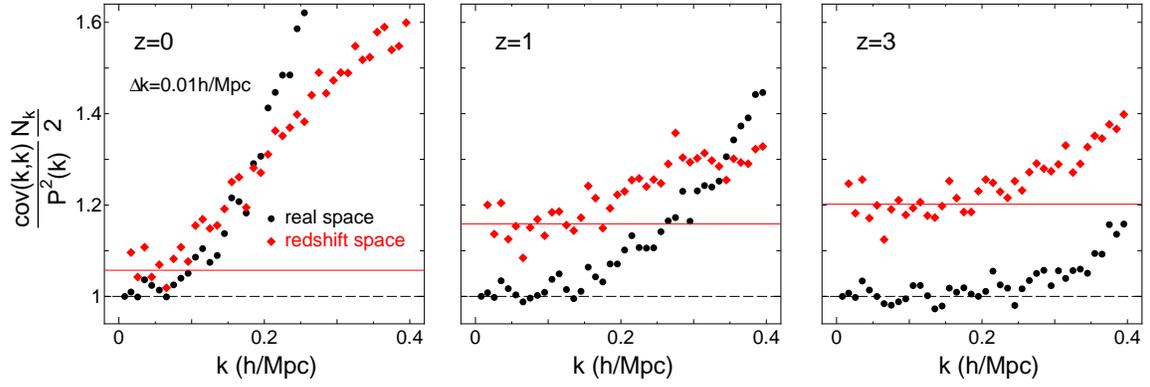}
\caption{
The diagonal covariance components as a function of wavenumbers, for
real- and redshift-space power spectra. 
Note that 
the redshift-space power spectrum studied here is the 
spherically averaged spectrum
over a shell of a given radius $k$ in redshift space 
(see text for the details). We show the covariances divided
by the Gaussian error contribution (the first term in Eq.~[\ref{cov}]):
 at large length scale limit ($k\rightarrow 0$), the real- and
redshift-space values approach to unity (solid line) and to the constant
factor that is given by the Kaiser's linear distortion
(dashed line), respectively.
The non-Gaussian error contribution is relatively suppressed in redshift
 space due to 
the nonlinear redshift distortions.
 } \label{pk_disp_redshift}
\end{figure*}

\begin{figure*}
\plotone{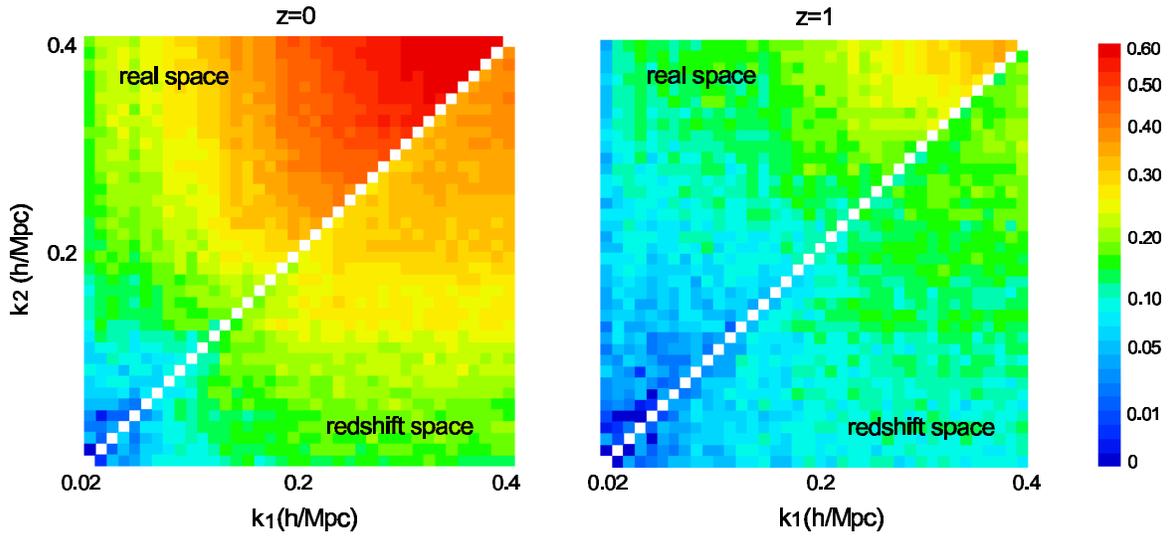}
\caption{The correlation matrix at $z=0$ (left panel) and $z=1$
(right panel).
In each panel, the upper-left matrix elements are the 
 off-diagonal covariances in  
real space, while the lower-right elements are for redshift
 space. 
}
\label{corrmat2}
\end{figure*}

The diamond symbols in Fig.\ref{pk_disp_redshift} show the diagonal
components of the redshift-space power spectrum covariances as a
function of wavenumbers and at three output redshifts. The diagonal
components are divided by the standard Gaussian errors, $2/N_k
P_{s0}^2(k)$, where we have used the {\it nonlinear} power spectrum
measured from the simulations.  Note the difference in the normalization
factor from that in Fig.\ref{pk_disp}. 
The horizontal line in each panel
shows the prefactor in Eq.~(\ref{disp_redshift2}), the amplification
factor expected from the Kaiser's formula at the large-scale
limit. Therefore the deviations from the horizontal line may come from
two contributions: (1) the non-Gaussian error contribution caused by
nonlinear clustering and (2) the nonlinear redshift distortions such as
the effect caused by the virial motions within and among halos, known as
the finger-of-God effect. The circle symbols denote the simulation
results for the real-space spectrum computed in a consistent way,
i.e. divided by the nonlinear spectrum. The nonlinear effects on the
covariance become more significant with increasing wavenumber and at
lower redshifts. Interestingly, however, comparing the real- and
redshift-space results manifests that 
the relative importance of the non-Gaussian covariances is
weaker in redshift space, implying that the finger-of-God
redshift distortions at small length scales more preferentially suppress
the covariance amplitudes than the power spectrum amplitudes (also see
Meiksin \& White 1999).

Fig.\ref{corrmat2} shows both the off-diagonal components of the covariances
in real space (the left-upper elements in each panel)
and redshift space (right-lower),  at redshifts $z=0$ and $z=1$.
The cross-correlations are more significant with increasing
wavenumbers, while the correlation strengths 
are relatively weaker in redshift space.

\section{Probability Distribution of the Power Spectrum Estimator}

\begin{figure}
\plotone{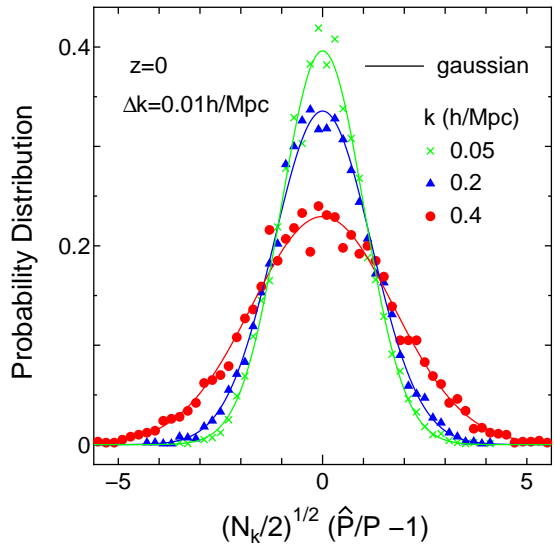}
\caption{Probability distribution of the power spectrum estimators
 $\hat{P}$ in the 5000 realizations.
The solid curves show the Gaussian distribution with zero mean and
variance that is set to the diagonal covariance component measured from
the simulations at a given wavelength.
}
\label{fig_pk_pdf}
\hspace{0.5cm}
\end{figure}

\begin{figure}
\plotone{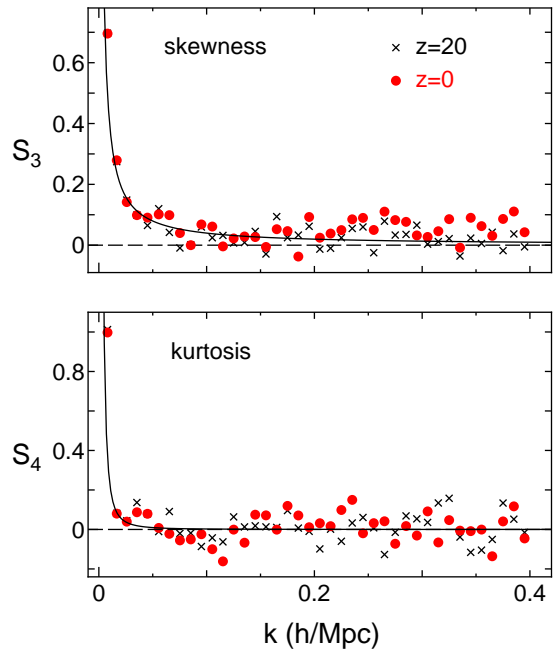}
\caption{The skewness (top panel) and the kurtosis (bottom panel) of the
 power spectrum distribution shown in the previous figure,  
 as a function of $k$ for outputs at $z=20$ and $0$.
The circles and crosses are the simulation results, while the solid curves
 are the theoretical predictions expected when 
the power spectrum estimators 
obey the $\chi^2$-distribution.
}
\label{fig_pk_skew_kurt}
\hspace{0.5cm}
\end{figure}

We have so far discussed the non-Gaussian covariance of the power
spectrum estimator $\hat{P}$.  It would be also intriguing to study how
the nonlinear clustering causes a non-Gaussian distribution in the power
spectrum estimators of a given $k$ among our 5000 realizations.  
For example, if the
estimators have a skewed distribution, a prior knowledge on the full
distribution may be needed to obtain an unbiased estimate on the
ensemble averaged band power at each $k$ from a small number of
realizations or a finite volume survey. Note that the power spectrum
covariance simply reflects the width (variance) of the full distribution
at each $k$'s but does not contain full information on the 
probability distribution.

Fig.~\ref{fig_pk_pdf} shows the probability distribution of the
power spectrum estimators $\hat{P}$ among 5000 realizations,
where we mean by ``probability'' that the distribution is normalized so
as to give unity if the distribution is integrated over the $x$-axis
values (see below).
The cross,
triangle, and circle symbols show the results for $k=0.05, 0.2,$ and $0.4
h/$Mpc, respectively.
The distribution is plotted as a function of $(N_k/2)^{1/2}
(\hat{P}/\bar{P}-1)$ for each $k$ such that the mean and variance of the
distribution are equal to zero and unity when the power spectrum 
distribution obeys
the linear-regime Gaussian distribution.  The simulation results show
that the distribution is broadened with increasing $k$ due to the stronger
nonlinearities. 
The solid curves show the {\em expected} Gaussian
distribution where its variance is set to the diagonal 
covariance measured from the simulations 
at each $k$, i.e. the variance includes the
non-Gaussian covariance contribution as given in
Fig.~\ref{pk_disp_redshift}. Interestingly, 
the simulation results are rather well approximated by the
 Gaussian distribution even in the non-linear regime.

The remaining small deviations from the 
Gaussian distribution
can be quantified by studying the skewness $S_3$ and kurtosis $S_4$
defined as
\beqa
  S_3=\frac{\langle ( {\hat P}(k)-P(k) )^3 \rangle}
  {\langle ( {\hat P}(k)-P(k) )^2 \rangle^{3/2}},  \nonumber \\
  S_4=\frac{\langle ( {\hat P}(k)-P(k) )^4 \rangle}
  {\langle ( {\hat P}(k)-P(k) )^2 \rangle^2}-3.
\label{s_3_4}
\eeqa
The $S_{3}$ and $S_4$ are vanishing for the Gaussian distribution. 
If the density field obeys the random Gaussian fields, which is a good
approximation in the linear regime, 
the power
spectrum estimator of a given $k$ (see Eq.~[\ref{eq_pk}]) obeys the
$\chi^2_{N_k}$-distribution in analogy with the CMB power spectrum (Knox
1995). 
In this case, as derived in Appendix B, 
the skewness and kurtosis can be analytically
computed as
\beq
  S_3=\sqrt{\frac{8}{N_k}},~~ S_4=\frac{12}{N_k}.
\label{eq_skew_kurt}
\eeq

Fig. \ref{fig_pk_skew_kurt} shows the simulation results for 
$S_{3}$ and $S_4$ as a function
 of $k$ at $z=20$ and $0$. 
The results are for a volume of $V=1h^{-3} {\rm Gpc}^3$, and the $S_{3,4}$
 scale as $S_3 \propto V^{-1/2}$ and $S_4 \propto V^{-1}$ from
 Eq.(\ref{eq_skew_kurt}).
The solid curves are the theoretical predictions of Eq.(\ref{eq_skew_kurt})
 which well match the simulation results. 
Note that the skewness is positive, 
 because the 
$\chi^2$-distribution has
 a long tail at large ends of $\hat{P}$.
Both $S_3$ and $S_4$ asymptotes to zero at high $k$, i.e., 
the probability distribution approaches to a Gaussian distribution 
at high $k$.
The skewness grows from $z=20$ to $0$ through the non-linear gravitational
 evolution, however, its value ($S_3 \lesssim 0.1$) is very small.

\section{Effects of Non-Gaussian Covariance on Signal-to-Noise Ratio}

A useful way to quantify the impact of the non-Gaussian errors is to
study the cumulative signal-to-noise ratio $(S/N)$ for measuring the
power spectrum over a range of wavenumbers, which is also sometimes 
called the 
Fisher information content (e.g. Tegmark et al. 1997).
The $S/N$ is defined, using the covariance (\ref{cov}), as
\beq
  \left( \frac{S}{N} \right)^2 = \sum_{k_{1}, k_2<k_{\rm max}}
P(k_1)  {\rm cov}^{-1} (k_1,k_2) 
P(k_2),
\label{sn}
\eeq
where ${\rm cov}^{-1}$ denotes the inverse of the covariance matrix and 
the summation is up to a given maximum wavenumber $k_{\rm max}$. 
Note that the $S/N$ is independent of the bin width assumed,
as long as the power spectrum does not vary rapidly within the bin widths.

\begin{figure*}
\epsscale{1}
\plotone{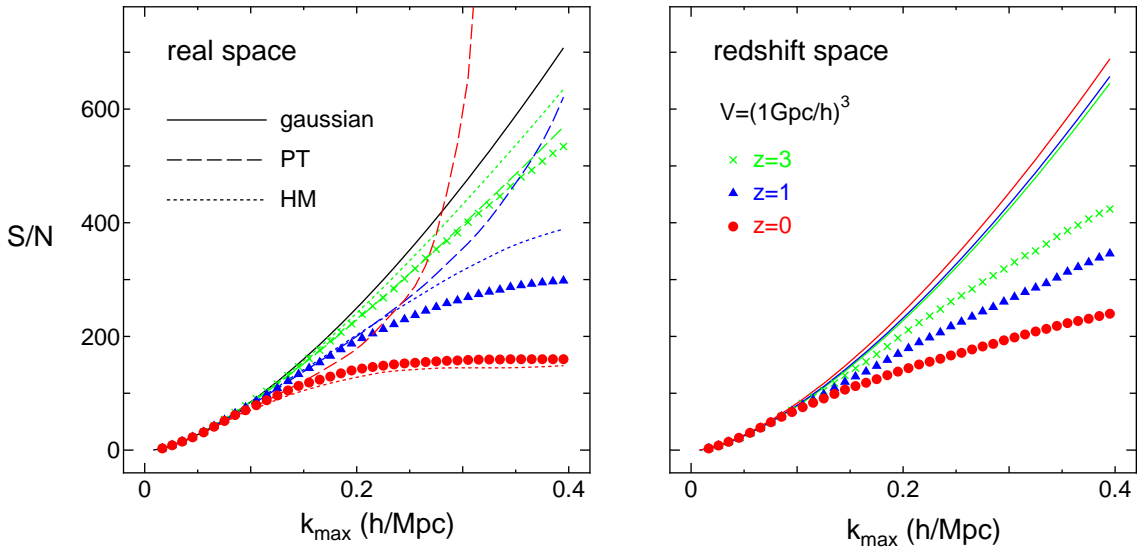}
\caption{The cumulative signal-to-noise ratios are shown 
as a function of $k_{\rm max}$ in real
space (left) and in redshift space (right), where the power spectrum 
information over
$2\pi/L\le k\le k_{\rm max}$ is included ($L$ is the box size).
Note that the $S/N$ amplitudes are for the simulation box volume
$V=1 h^{-3}$Gpc$^{3}$. 
The solid curves in each panel
show the $S/N$ for the Gaussian covariance case, which scales as 
$S/N\propto k_{\rm max}^{3/2}$. 
The simulation results increasingly deviate from the Gaussian error
 results with increasing $k_{\rm max}$.
In the left panel the dashed curve shows the analytical prediction for
$S/N$ when the PT trispectrum is used to model the non-Gaussian
covariance (the simulation result is used for the power spectrum in the
$S/N$ calculation). The dotted curves show the results when the halo
 model is used to model the non-Gaussian covariance. 
} \label{fig_sn}
\vspace*{.5cm}
\end{figure*}

Fig.\ref{fig_sn} shows the $S/N$ as a
 function of $k_{\rm max}$ for the spectra 
at $z=0,1$ and $3$ in real space (left panel) and in 
redshift space (right), respectively. The results for $S/N$ shown here 
are for a volume of $V=1h^{-3}{\rm Gpc}^3$
(the $S/N$ scales with $V$ as $S/N \propto V^{1/2} $).
For comparison the solid curve shows the $S/N$ for the Gaussian error 
case, which scales as $S/N\propto k_{\rm max}^{3/2}$ independently of
redshift. 
The simulation results show that the non-Gaussian errors degrade the
$S/N$. The degradation becomes more significant with increasing $k_{\rm
max}$ and at lower redshifts: for the results in real space the $S/N$ is
degraded by up to a factor of $4$ and $2$ for $z=0$ and $1$, respectively,
compared to the Gaussian error case.  It should be worth noting that the
$S/N$ becomes nearly constant on $k_{\rm max}\simgt 0.2$ and
0.3$h$Mpc$^{-1}$, i.e no gain in the $S/N$ even if including modes
at the larger $k$, as has been found in the previous works (Rimes \&
Hamilton 2005, 2006; Hamilton, Rimes \& Scoccimarro 2006; Neyrinck,
Szapudi \& Rimes 2006; Neyrinck \& Szapudi 2007; Lee \& Pen 2008;
Angulo et al. 2008; Smith 2008).

In the dashed curves we use the perturbation theory (PT) in 
Eq.~(\ref{eq:pt}) to calculate the covariance, while the power spectrum
measured from the simulations is used for the numerator in the $S/N$
calculation. 
The PT provides a better fit to the data in the linear and weakly
 nonlinear regime, which is coincident within $10\%$ for $k_{\rm
 max}<0.16h/$Mpc at $z=0$, $k_{\rm max}<0.23h/$Mpc at $z=1$ and $k_{\rm
 max}<0.4h/$Mpc at $z=3$, respectively.
However, at small scale the deviation is so large, since the theory predicts
 the much smaller covariance than the data as shown in
 Figs.~\ref{pk_disp} and \ref{corrmat}.

The dotted curves in the left panel of Fig.~\ref{fig_sn} show the halo
model results, which fairly well fit the simulation results at $z=0$.
At higher redshifts ($z=1,3$), the halo model reproduces a saturation in the
$S/N$ amplitude on the small $k_{\rm max}$, but underestimates the
impact of the non-Gaussian errors, which is due to the underestimation
in the off-diagonal elements of the covariances in the halo model (see
Fig.~\ref{corrmat}).

The impact of the non-Gaussian errors on $S/N$ is mitigated in redshift
space as in Fig.~\ref{pk_disp_redshift}. Also note that the $S/N$ in
redshift space continues to increase 
for the smaller scales $k_{\rm max}>0.2h/$Mpc. This is again because the
nonlinear redshift distortions cause strong suppression in the
non-Gaussian covariances, making the $S/N$
closer to the Gaussian error case. 

\begin{deluxetable}{cccc}
\tablecaption{WFMOS Survey Parameters}
\tablewidth{0pt}
\tablehead{
  Redshift & Volume & Number Density 
 & Bias \\
 &  ($h^{-3}$Gpc$^3$) & ($h^{-3}$Mpc$^{-3}$)
 & 
}
\startdata
 $0.5-1.3$ & $4.0$ & $5 \times 10^{-4}$ 
 & $1.7$ \\
 $2.3-3.3$ & $1.0 $ & $5 \times 10^{-4}$
 & $3.2$
\enddata
\label{table1}
\end{deluxetable}

\begin{figure}
\plotone{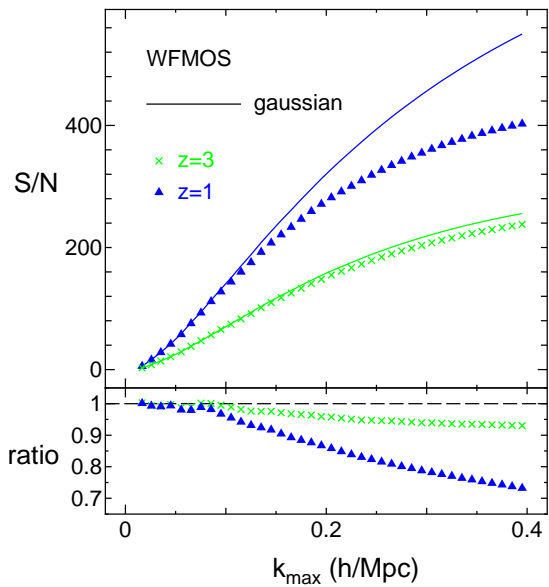}
\caption{The expected $S/N$ for 
 model WFMOS surveys of $z \sim 1$ and $z \sim 3$ slices 
(see Table~\ref{table1}). Note that the shot noise contribution to the 
covariance 
is included.
The solid curves denote the $S/N$ without the non-Gaussian covariance
 contribution.
The bottom panel shows the ratio between the simulation result and 
the solid curve for each redshift slice.
The Gaussian error assumption overestimates the $S/N$ by $30\%$
 ($7\%$) at $z=1(3)$ for $k_{\rm max}=0.4h/$Mpc.
}
\label{fig_sn_wfmos}
\hspace{0.5cm}
\end{figure}

We make a more realistic estimate for the $S/N$ 
taking into account the shot noise effect of galaxies that are biased
tracers of large-scale structure. To do this we consider a galaxy survey
that resembles the planned survey by 
WFMOS (Wide-Field Fiber-Fed Optical Multi-Object Spectrograph), 
and assume the fiducial survey parameters given in Table~\ref{table1}
(see also the WFMOS feasibility report\footnote{Feasibility Study Report :
 http://www.gemini.edu/files
 /docman/science/aspen/WFMOS$\_$feasibility$\_$report$\_$public.pdf}).
The target galaxies are supposed to be emission line galaxies and
Lyman-break galaxies, at $z\sim 1$ and $z\sim 3$, respectively.
To compute the power spectrum and covariance of galaxies, we employ a
linear bias model for simplicity.
This assumption is not accurate since the bias is generally nonlinear and
 scale-dependent (Smith et al. 2007), and hence our results below 
just give a rough
 estimate on the $S/N$.
The bias parameters in Table~\ref{table1} are chosen such that the 
rms density fluctuations of galaxies within a sphere of radius 
$8h^{-1}$Mpc
become $\sigma_{g8}=0.8$ (Glazebrook et al. 2005).
We simply include the effects of the linear bias and the shot noise
 by replacing the power spectrum and the covariance in the $S/N$
 evaluation as $P\rightarrow b^2P$ and 
${\rm cov}\rightarrow b^4{\rm cov}+2b^2P\bar{n}_g^{-1}+\bar{n}_g^{-2}$,
where $\bar{n}_g$ is the mean number density of galaxies. 
The above replacement is done in real space.

The symbols in Fig. \ref{fig_sn_wfmos} show the simulation results 
in real space.  The expected $S/N$ is found
to be very significant: $S/N\simeq 400$ and 200 for the slices of $z=1$
and $3$ for $k_{\rm max}=0.4h/$Mpc, respectively. This implies that the
WFMOS-type survey 
allows a precision of measuring the power spectrum amplitudes at a
sub-percent level.\footnote{The Fisher information matrix for the power
spectrum measurement is given as 
$F_{\mi \mj}=\sum_{k_{1,2}} [\partial P(k_1)/\partial \theta_\mi] ~{\rm
cov}^{-1}(k_1,k_2) [\partial P(k_2)/\partial \theta_\mj]$, where
$\theta_\mi$ is a set of cosmological parameters of interest.  Roughly
speaking, the {\em unmarzinalized} uncertainty in estimating a parameter
$\theta_\mi$ is given as $\sigma^2(\theta_i)=[F_{ii}]^{-1/2}\propto
(S/N)^{-1}$ (this exactly holds if the power spectrum amplitude, for
the fixed shape, is considered as the parameter). Therefore the $S/N$
amplitude gives a rough estimate on the precision of parameter
estimation provided the power spectrum measurement: the greater $S/N$
means the higher precision.}
The solid curves
are the results obtained assuming the Gaussian covariances with the shot
noise contribution, which do not scale as $S/N\propto k_{\rm max}^{3/2}
$ on scales where the shot noise is relevant in the covariance
($\bar{n}_gP\simgt 1$).  Compared with Fig.~\ref{fig_sn}, one can find
that the shot noise causes positive and negative effects on $S/N$: it
reduces the overall amplitudes of $S/N$, but mitigates the degradation
due to the non-Gaussian errors.

Since the precision of constraining individual cosmological parameters
such as dark energy parameters roughly scales with the $S/N$
 amplitude\footnotemark[6],
Fig.~\ref{fig_sn_wfmos} implies that the constraining power of the
fiducial WFMOS survey is degraded by the non-Gaussian
errors, compared with the Gaussian error case.
The impact of the non-Gaussian errors on cosmological parameter estimations
 will be presented in a subsequent paper (Takahashi et al. in preparation).

\section{Effects of Long-Wavelength Fluctuations}
\label{sec:period}

We have so far employed, as usual, the simulations with periodic
boundary conditions, where there is no clustering power on scales greater
than the simulation box. Obviously, however, the real universe never
obeys the periodic boundary condition and does contain the density
perturbations of scales greater than a surveyed volume.  In particular,
Rimes \& Hamilton (2006) pointed out a new source of the non-Gaussian
errors that inevitably arises when the power spectrum is estimated from a
finite-size volume, called the beat-coupling (BC) effect (2006; also see
Hamilton, Rimes \& Scoccimarro 2006; Sefusatti et al. 2006). If the
survey region is embedded in a large-scale overdensity or underdensity
region, then the small scale fluctuations we measure may have
grown more rapidly or slowly than the ensemble average. There are thus
non-vanishing correlations of the small-scale fluctuations with the
unseen large-scale fluctuations.  This physical correlations may add
uncertainties in measuring the power spectrum on scales of interest.

In this section, therefore, we study how the periodic boundary
conditions and the density perturbations larger than a survey volume
(the volume where the Fourier transform is performed) affect the power
spectrum estimation and the covariance. For this purpose we study the
following three cases:
\begin{itemize}
\item[Case 1:] We first divide each simulation region of $1h^{-3}$Gpc$^3$
      into eight cubic sub-boxes of equal volume.
      Each sub-box has a volume of $(500h^{-1}$Mpc$)^3$ and contains
      about $128^3$ particles. We then randomly select only one sub-box
	   and use the particle distribution to resemble the density
	     perturbation field. The density perturbation field
	     outside the sub-box is zero-padded within the whole
      box of $1h^{-3}$Gpc$^3$. The mean mass density 
is computed from the number of the particles within the sub-box.
Then we perform the FFT of 512$^3$
      grids for the whole box to estimate the power spectrum. 

\item[Case 2:] Similar to Case 1, but the FFT of 256$^3$ grids is
	     performed only
      within the sub-box of volume ($500h^{-1}$Mpc$)^3$ that contains
      the    $N$-body particles (therefore no zero-padded
      region).

\item[Case 3:] We run new simulations of volume $(500h^{-1}$Mpc$)^3$
	     using $128^3$
      particles and employing the periodic boundary condition. Then the
      power spectrum is estimated from the whole box using the FFT of
      256$^3$ grids. 
\end{itemize}
Note that the effective mass and spatial resolutions are the same in
all the cases.
We use 400 realizations for each case. Cases 1 and 2 do not employ
the periodic boundary condition and contain the density fluctuations
larger than the FFT-used volume in structure formation. However, these
two cases are different in that the fundamental mode of Fourier
transform, given as $\epsilon\equiv 2\pi/L$ ($L$ is the size of FFT
volume), is smaller in Case 1 by factor 2 than in Case 2. Therefore the
density fluctuation field is sampled by the finner Fourier modes in Case
1. Also note that Case 1 corresponds to a case that the FFT transform is
applied to a survey with complex geometry.  Case 3 has the periodic
boundary condition, and is equivalent to the procedure we have employed
up to the preceding section. By comparing these three cases, we will
below address the effects of the periodic boundary condition, the finite
Fourier sampling and the beat-coupling effect.

\begin{figure}
\epsscale{1.}
\plotone{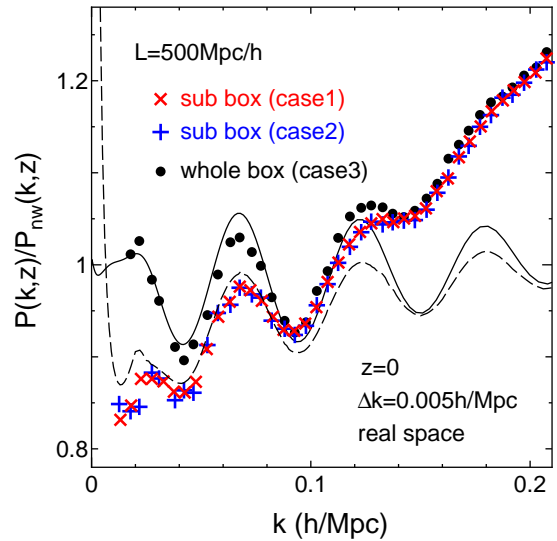}
\caption{
The real-space power spectra at $z=0$, computed based on the procedures
of Cases 1, 2 and 3 in \S~\ref{sec:period}. Note that the power spectra
are divided by the non-wiggle linear power spectrum of Eisenstein \& Hu
(1999) for illustrative purpose.  The dotted symbols show the spectrum
estimated from the simulations with the periodic boundary condition.
The cross and plus symbols show the results without the periodic
boundary condition: the results include contributions from
the density perturbations of scales greater than the Fourier-transformed
 volume. 
The density perturbation fields for the cross and
plus symbols are equivalent, but
the Fourier-transformed volume for the cross symbol is set to contain
the zero-padded region (see text for the details). 
The solid curve denotes the linear theory prediction.
The dashed curve is the same as the solid curve, but convolved with
 the window function. 
}
\label{fig_pk_mean}
\hspace{0.5cm}
\end{figure}

Fig.~\ref{fig_pk_mean} shows the real-space power spectra at $z=0$,
estimated according to the procedures described above, where the
simulated power spectra are divided by the non-wiggle linear power
spectrum in Eisenstein \& Hu (1999) for illustrative purpose. 
The cross, plus and dotted symbols are the results for
Cases 1, 2 and 3, respectively. 
All the results agree well on scales $k\simgt 0.1h$Mpc$^{-1}$. However,
the results for Cases 1 and 2, which do not impose the periodic boundary
condition, underestimate the power spectrum amplitudes at the linear
regime $k\sim 0.1h$Mpc$^{-1}$ by up to $10\%$. This is because the
non-periodic density fluctuation field is expanded by the FFT transform
that has periodic basis eigenfunctions within the box size (see Sirko
2005 for the similar discussion).  This underestimation can be corrected
for if the Fourier kernel of the non-periodic field is properly taken
into account.
The dashed curve is the linear power spectrum convolved with the 
 window function which is given as 
$W(\bfx)=1$ $(=0)$ inside (outside) the sub box:
\beq
  P_W(\bfk)=\frac{1}{V} \int \frac{d^3 k^\prime}{(2 \pi)^3} 
  P(k^\prime) \left| \tilde{W}(\bfk-\bfk^\prime) \right|^2,
\eeq
The window function in Fourier space is
\beq
 \tilde{W}(\bfk) = V
 \prod_{\mi=x,y,z} \frac{\sin(k_\mi L/2)}{k_\mi L/2},
\eeq 
with $L=500 h^{-1}$Mpc.
The dashed curve is the spherical averaged power spectrum,
 $P_{W0}(k) = \int d\Omega_k/(4 \pi) ~P_W(\bfk)$, which
 reproduces the dumping of the power spectrum
 at $k<0.1h/$Mpc.

\begin{figure*}
\epsscale{0.8} \plotone{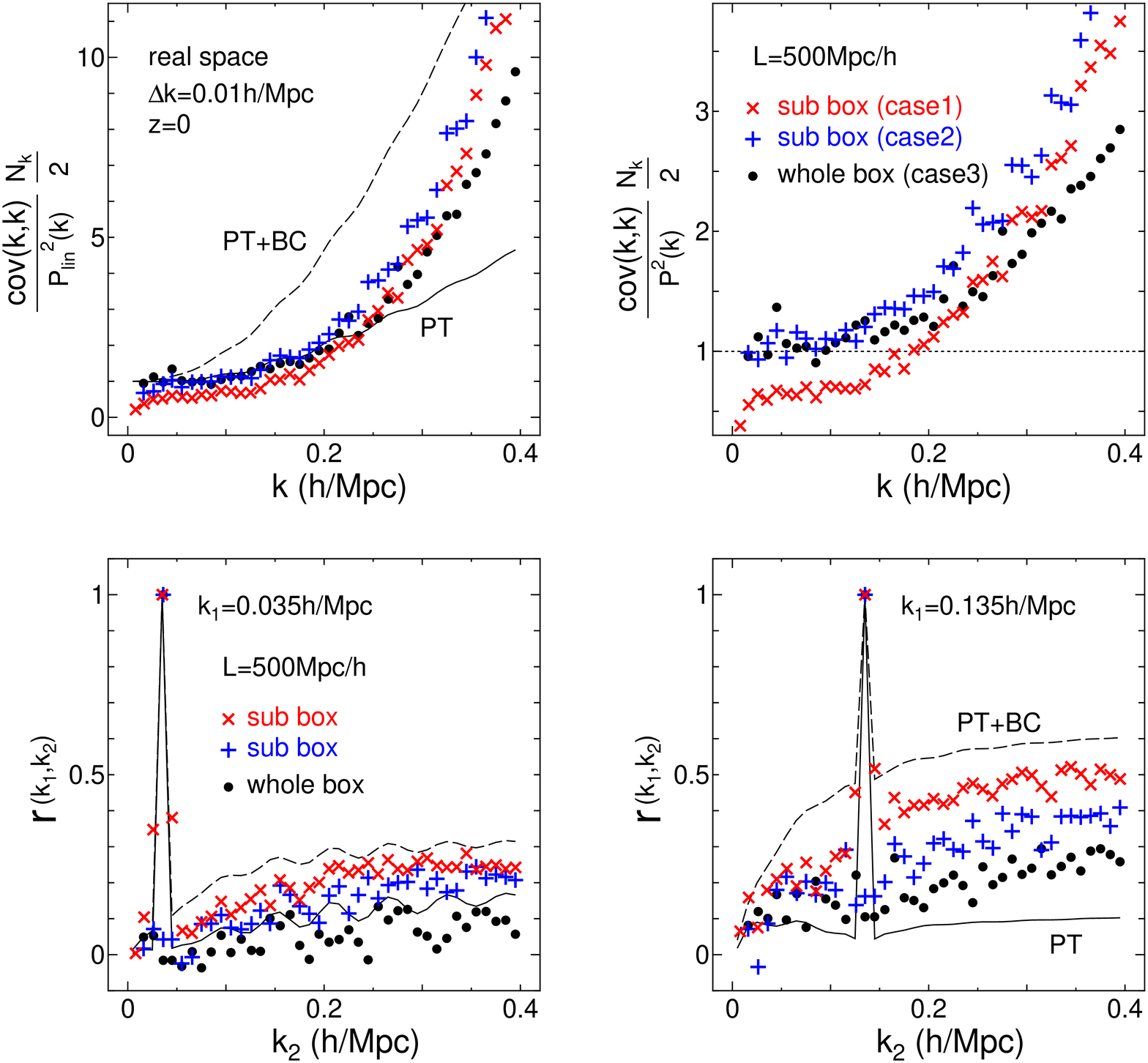} 
\caption{The power spectrum covariances in real space at $z=0$, computed
from the three numerical settings as in Fig.~\ref{fig_pk_mean}.  The
diagonal components of the covariances as a function of $k$, normalized
by the linear Gaussian covariances (upper-left panel) and by the
 measured Gaussian covariances (upper-right), respectively. 
The bottom panels show the correlation coefficients $r(k_1,k_2)$ for the
 covariances as a function of $k_2$, for $k_1=0.035h$Mpc$^{-1}$ (left)
 and 0.135$h$Mpc$^{-1}$ (right), respectively. In all the panels, the
 solid curves show the analytical predictions for $z=0$ obtained when the PT
 is used to model the non-Gaussian covariances.
 The dashed curves show the results when the additional
 non-Gaussian errors due to the long-wavelength fluctuations, modeled by 
Eq.(\ref{bc}), are
 further included.
}
\label{fig_cov_bc}
\hspace{0.5cm}
\end{figure*}

\begin{figure*}
\epsscale{1.1}
\plotone{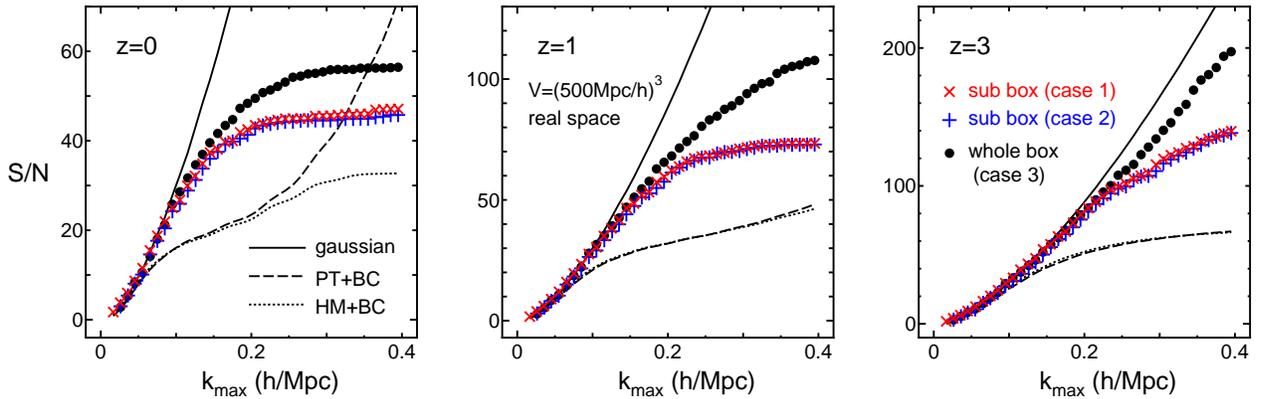}
\caption{The signal-to-noise ratios at redshifts $z=0$ (left), $1$
 (center) and 3 (right),
 computed from the three numerical settings as in Fig.~\ref{fig_pk_mean}. 
The dotted symbols show the equivalent results to Fig.~\ref{fig_sn}.
The cross and plus symbols show the results, where the simulations
without the periodic boundary condition are used. These simulations
include contributions from the fluctuations at length scales greater 
than the Fourier volume -- the beat-coupling (BC) effect.
The solid curves are the result for the Gaussian error case, while
the dashed curves show the results of three redshifts
obtained when the non-Gaussian covariances are computed from the PT
trispectrum plus the BC effect modeled by Eq.~(\ref{bc}). The naive BC
model significantly overestimates the non-Gaussian errors seen from the
simulations. The dotted curves show the result obtained when the halo
model trispectrum is further included.
} \label{fig_sn_bc} \hspace{0.5cm}
\end{figure*}
Fig.\ref{fig_cov_bc} compares the power spectrum covariances.
The top two panels show the diagonal parts normalized by the linear
power spectrum (upper-left) and the spectrum measured from simulations
(upper-right), which are similar to Figs.\ref{pk_disp} and
\ref{pk_disp_redshift}, respectively.
The number of modes $N_k$ in the vertical axis is used for the
 $(500 {\rm Mpc}/h)^3$ box for all the cases.
First, comparing the results for Cases 2 and 3, we find that there are stronger
non-Gaussian errors at $k\simgt 0.2h$Mpc$^{-1}$ for Case 2: the presence
of the density fluctuations larger than the survey volume, as in the real
universe, increases the non-Gaussian error strengths due to the
mode-coupling between the large- and small-scale density
fluctuations. Next, let's compare Case 1 with Cases 2 and 3.  The
results for Case 1 are clearly smaller than Cases 2 and 3 even at the
linear scales such as $k\simlt 0.1h$Mpc$^{-1}$, but show stronger
non-Gaussian errors than Case 3 on the large $k$'s as in Case 2.  The
differences between Cases 1 and 2 are caused by the presence of the zero-padded
regions within the FFT volume and the finer Fourier sampling. For Case
1, the density perturbations of scales comparable to the FFT volume
are non-periodic due to a mixture of the zero-padded region and the
N-body particle distribution in the sub-volume. Therefore, the Fourier
transform causes artificial cross-correlations between the
Fourier modes of different $\bf k$'s even in the
linear regime. Due to the cross-correlations, the off-diagonal
covariances are amplified as shown in the lower panels, while the
diagonal covariances are relatively suppressed.

The behaviors of the off-diagonal covariances shown in the lower panels
are similarly understood. Comparing the three cases one can find that
the long-wavelength fluctuation effect and the zero-padding plus the
finer Fourier sampling cause stronger cross-correlations between the
spectra of different $k$'s over the range of $k$ we have considered.

More important results are given in Figs.~\ref{fig_sn_bc}, showing the
cumulative $S/N$ values for Cases 1, 2 and 3,
computed properly taking into account the covariances in
Figs.~\ref{fig_cov_bc}.
Each panel shows the results at redshifts of $z=0$ (left), $1$ (center)
 and $3$ (right).
First of all, all the results agree well with
the Gaussian error case on the linear scales, $k\simlt 0.1 h$Mpc$^{-1}$.
At the larger $k$'s, the results for Cases 1 and 2 show that the
presence of long-wavelength fluctuations 
further degrades the $S/N$ amplitudes by 20\% at
redshift $z=0$ and by 30\% at $z=1$ and 3, respectively, compared to the
results with the periodic boundary condition (Case 3). This implies
that, even for high redshifts and at the BAO scales, the additional
non-Gaussian errors due to the long-wavelength fluctuation 
effect need to be included
in the analysis for an actual survey.
Interestingly, the $S/N$ values for Cases 1 and 2 become to agree well, even
though their covariances are very different as shown in
Fig.~\ref{fig_cov_bc}. This agreement is reasonable, because Cases 1 and
2 contain the similar mass density fields of same volume; therefore
the amount of cosmological information to be extracted is similar.
These are different only in the FFT procedures.

A full physical understanding of the complex covariance behaviors is
beyond the scope of this paper. Nevertheless, it would be interesting to
compare the simulation results with an analytical model.  A crude model
to describe the long-wavelength fluctuation effect on the
non-Gaussian covariance is proposed in Hamilton et al. based on the
perturbation theory (also see Takada \& Jain 2008):
\beq
  {\rm cov}_{\rm BC}(k_1,k_2) = \frac{1}{V} 16
 \left( \frac{17}{21} \right)^2 P_{\rm lin}(\epsilon) 
 P_{\rm lin}(k_1) P_{\rm lin}(k_2),
\label{bc} 
\eeq
where $\epsilon \equiv \pi/L$ ($L=500h$Mpc$^{-1}$ for Cases 1 and 2).
This model ignores the Fourier transform effect of the non-periodic
density field and rests on a simplified assumption that the
long-wavelength fluctuation effect arises from a correlation of the
fundamental Fourier mode $\epsilon$ with the wavenumbers we want to
measure.
Developing a
more accurate analytical model of the long-wavelength fluctuation 
effect is now in
progress and will be presented elsewhere (Kayo et al. in preparation).

The solid curves in Fig.~\ref{fig_cov_bc} show the $S/N$
for the Gaussian error case, 
and roughly explains the simulation results
up to the linear regimes. The dashed curves
show the PT  model predictions including the BC effect modeled by 
Eq.~(\ref{bc}), 
which are intended to reproduce the results for Case
2. The dotted curves show the results obtained when the halo model
contribution to the covariance is further included.
These analytic models do not describe the complex behaviors of the
diagonal and off-diagonal covariances seen in the simulations. Also the
simulation results for $S/N$ cannot be explained by the analytic
models.
The analytical model in Eq.(\ref{bc}) overestimates the beat-coupling
 effect. This conclusion agrees with a recent study of Reid, Spergel
 \& Bode (2008), where 
they showed that the simulation results are fairly well explained if 
Eq.~(\ref{bc}) reduced by a factor 3 is added to the Gaussian
covariance. 

\section{Discussion and Conclusion}

Having well-calibrated, accurate covariances of 
the power spectrum is clearly needed in order to obtain unbiased, robust
cosmological constraints from ongoing/future BAO experiments. 
In previous studies, the covariance matrix is calculated either by
using analytic models which cannot be applied to fully nonlinear regimes,
or by using a limited
 number of simulation realizations  (Scoccimarro et al.
 1999; Meiksin \& White 1999; Rimes \& Hamilton 2005 and Neyrinck \& Szapudi 2008). 
In this paper, we used a very large number ($5000$) of the realizations
to study the power spectrum covariances, allowing us to achieve 
the convergence at a few \% level. 

We have carefully studied how the non-Gaussian error contributions to
the covariance vary with scales and redshifts for the concordance
$\Lambda$CDM model. As expected in the CDM model, the non-Gaussian
errors become more significant on smaller length scales and at lower
redshifts. For redshifts $z=0$, 1 and $3$, the cumulative
signal-to-noise ($S/N$) ratios for measuring the power spectrum over
$0.01\simlt k \simlt 0.4h$Mpc$^{-1}$ are degraded due to the
non-Gaussian errors by a factor of 1.3, 2.3 and 4, respectively,
compared to the Gaussian error cases. This degradation is slightly
mitigated in redshift space because the nonlinear redshift distortions
cause a stronger suppression in the covariance amplitudes than in
the power spectrum amplitudes.

We also estimated how the density fluctuations of scales greater
than a survey size cause additional non-Gaussian errors via the
correlations with the fluctuations we want to measure, which inevitably
arises for a finite-size survey -- the so-called beat-coupling effect.
This effect disappears when estimating the power spectrum covariances
from simulations with the periodic boundary condition. Thus we
rather used the sub-region of the original simulation to estimate the
new non-Gaussian errors, and showed that the beat coupling effect can be
important even in the weakly nonlinear regime and for high redshifts: it
further suppresses the $S/N$ by $20 \%$ at $z=0$, and $30 \%$ at
$z=1$ and $3$, respectively.  However, the behaviors of these non-Gaussian
errors cannot be described by the naive analytic models with and without 
the beat-coupling effect. 
Therefore it will be worth exploring a more accurate analytical model of the
non-Gaussian covariances. Such a model will help us to obtain physical
interpretation and to calibrate the derived covariance for
arbitrary cosmological models and survey parameters (Kayo et al. in
preparation).

We also studied the probability distribution of the power spectrum
estimators among the 5000 realizations. We found that the distribution
is nearly Gaussian even in the nonlinear regime. More precisely, the
mean of the power spectrum estimators is not largely
biased from the ensemble average, 
and the scatters are well given by the diagonal 
power spectrum covariance at a given wavenumber.

A more important question would be how an actual galaxy survey is
affected by the non-Gaussian errors. For this purpose, 
we made a simplified estimate on the $S/N$ expected for a
WFMOS-type survey, further 
taking into account the shot noise contamination to
the covariance due to finite number densities of galaxies.
Since the shot noise contributes only to the Gaussian errors
(in an ideal case), including the shot noise not only reduces the
total $S/N$ amplitude, but also mitigates the influence of the
non-Gaussian errors. Thus the impact of the non-Gaussian errors does
vary with survey parameters. Since the precision of a given survey for
constraining cosmological parameters roughly scales with the $S/N$
amplitude, an optimal survey design needs to be realized by taking
into account the non-Gaussian errors, given the resources and observing
times for a survey.  Furthermore the non-Gaussian errors may cause the
best-fitting parameters to be biased if the model fitting is done
improperly assuming the Gaussian covariances, because the non-Gaussian
errors are more significant on smaller length scales and cause
correlated uncertainties between the band powers. These issues will 
be studied in a forthcoming paper (Takahashi et al. in preparation).

Our simulation results of the power spectrum $P(k)$ and
the covariance matrix ${\rm cov}(k_1,k_2)$ are available
as numeric tables upon request (contact takahasi@a.phys.nagoya-u.ac.jp).

\acknowledgments
We would like to thank Masanori Sato for useful comments and discussions. 
This work is supported in part by Grant-in-Aid for Scientific Research
on Priority Areas No. 467 ``Probing the Dark Energy through an
Extremely Wide and Deep Survey with Subaru Telescope'',
 by the Grand-in-Aid for the Global COE Program
 ``Quest for Fundamental Principles in the Universe: from Particles
 to the Solar System and the Cosmos'' from the Ministry of Education,
 Culture, Sports, Science and Technology (MEXT) of Japan, 
by the World Premier
International Research Center Initiative of MEXT of Japan,
 by the Mitsubishi Foundation,
and by Japan Society for Promotion of Science (JSPS) Core-to-Core
Program ``International Research Network for Dark Energy'', and by
Grant-in-Aids for Scientific Research
(Nos.~18740132,~18540277,~18654047).  I. K. and T. N. is
supported by Grants-in-Aid for Japan Society for the Promotion of
Science Fellows.

\appendix
\section{Convergence of the Covariance Matrix}

\begin{figure}
\plotone{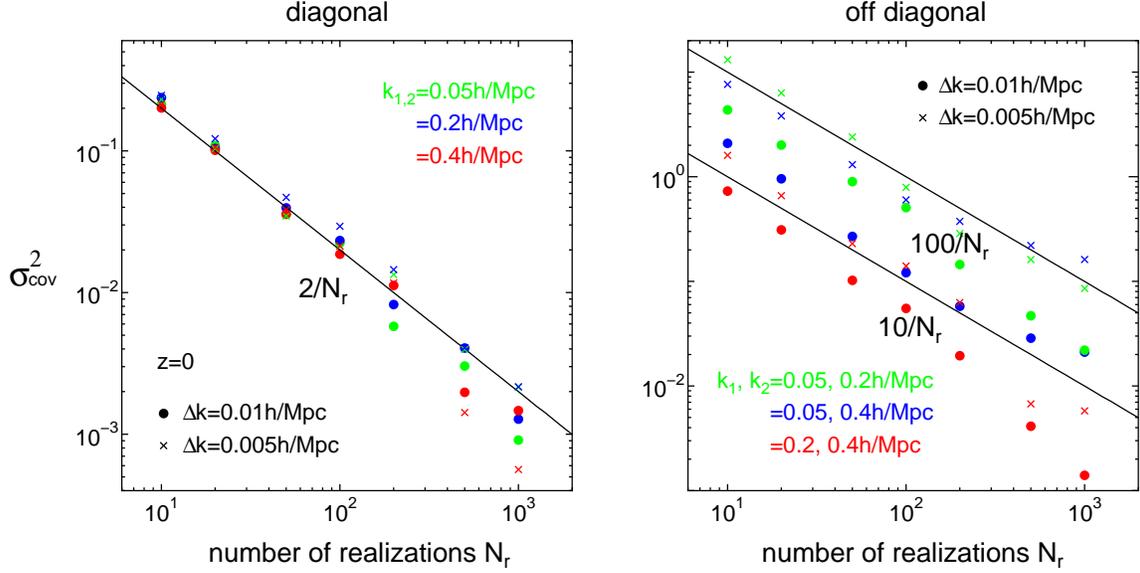} 
\caption{ The dispersions among the power spectrum
covariances each of which is estimated from the $N_{\rm r}$ realizations
(a subset of the while 5000 realizations), as a function of $N_{\rm r}$.
The dispersion is estimated using Eq.~(\ref{disp_cov}).
The left (right) panel show the results 
for the diagonal (off-diagonal) parts for varying the wavenumber bins
 and the bin widths.  The
color symbols are the simulation results, while the solid curves denote the
approximate fittings (see text for the details). The plots explicitly
show that the power spectrum covariances are estimated at a sub-percent
level accuracy by using our whole 5000 realizations.  }
\label{fig_convergence_cov} \hspace{0.5cm}
\end{figure}

It is useful to estimate the necessary number of realizations
to achieve a desired accuracy for estimating the power spectrum covariance.
In this appendix, we examine the numerical convergence of the covariance
estimation.
Let us define the dispersion of the covariance matrix as
\beq
  \sigma_{\rm{cov}}^2 (k_1,k_2) =
  \frac{\langle ({\rm{cov}}(k_1,k_2) - \langle {\rm cov}(k_1,k_2)
  \rangle )^2 \rangle}
  {\langle {\rm cov}(k_1,k_2) \rangle^2}.
\label{disp_cov}
\eeq
Here, $\langle \rm{cov}(k_1,k_2) \rangle$ is the ensemble average 
over all the realizations, while $\rm{cov}(k_1,k_2)$ is the covariance 
estimated from a subset of the realizations whose number is denoted as
 $N_{\rm{r}}$.
Fig.~\ref{fig_convergence_cov} shows the dispersion in Eq.(\ref{disp_cov}) 
 as a function of the number of realizations $N_{\rm{r}}$.
The left panel is for the diagonal elements,  and each color symbols correspond
 to $k_{1,2}=0.05$ (green), $0.2$ (blue) and $0.4h/$Mpc (red) with
 the bin width $\Delta k=0.01$ (circles) and $0.005h/$Mpc (crosses).
The solid line represents $2/N_{\rm{r}}$ which fits the data very well.
Hence, we numerically find the scaling of the dispersion given by
\beq
    \sigma_{\rm{cov}}^2 (k_1,k_1) \simeq \frac{2}{N_r}.
\label{disp_cov2}
\eeq
Note that the above result
 is independent of the scale $k$, the bin width $\Delta k$ and the
 simulation box volume.

The right panel is for the off-diagonal elements for
 $(k_1,k_2)=(0.05,0.2)$, $(0.05,0.4)$
and $(0.2,0.4)h/$Mpc, respectively.
The solid lines represent $10/N_{\rm{r}}$ and $100/N_{\rm{r}}$.
Similar to the diagonal parts,
 we obtain
\beq
    \sigma_{\rm{cov}}^2 (k_1,k_2) \propto \frac{1}{N_r} \frac{1}{(\Delta k)^2}
    \propto \frac{1}{N_r} \frac{1}{\Delta N_{k1} \Delta N_{k2}},
\label{disp_cov3}
\eeq
for $k_1 \neq k_2$.
Here $\Delta N_{k_i}$ ($i=1,2$) are the numbers of modes available for 
the bins $k_{i}$ with the bin width, 
and the proportionality factor would depend on the scale $k_{i}$.
The analytical derivation for Eqs.(\ref{disp_cov2}) and (\ref{disp_cov3})
 will be presented in Kayo et al. (in preparation).

\section{Probability Distribution of the Power Spectrum Estimator in Gaussian
 Limit}

In the linear regime, the real and imaginary parts of the density fluctuation
 (${\rm Re}[\delta_\bfk]$ and ${\rm Im}[\delta_\bfk]$)
 follow the Gaussian distribution with mean $0$ and dispersion $P(k)/2$.
The power spectrum estimator for a given realization $\hat{P}$ is the
 summation of the squared Gaussian fluctuations, $\hat{P}(k)=(1/N_k)
 \sum_\bfk ({\rm Re}[\delta_\bfk]^2+{\rm Im}[\delta_\bfk]^2$). 
Then the distribution of $\hat{P}$ obeys the chi-square distribution
 (e.g. Abramowitz \& Stegun 1970):
\beq
  F(\hat{P}(k);N_k/2)=\frac{1}{\Gamma(N_k/2)} \left( \frac{N_k}{2}
 \frac{\hat{P}(k)}{P(k)}
  e^{-\hat{P}(k)/P(k)} \right)^{N_k/2} \frac{1}{\hat{P}(k)}
\label{chi_sq_disp}
\eeq
for $\hat{P}>0$ and $F=0$ for $\hat{P}<0$.
Its mean and dispersion are $P$ and $P^2/(N_k/2)$, respectively.
The skewness and kurtosis are given in Eq.(\ref{eq_skew_kurt}).
In the limit of $N_k \rightarrow \infty$ it reduces to the Gaussian
 distribution.
The factor two in the number of modes $N_k/2$ arises because
 $\delta_\bfk$ and $\delta_{-\bfk}$ are not independent.

\clearpage

\end{document}